\documentclass{pasj00}
\draft

\usepackage{color}
\usepackage{ulem}

\begin{document}
\SetRunningHead{Y. Ezoe et al.}{
Suzaku and XMM-Newton Observations of the Eastern Tip 
Region of the Carina Nebula}
\Received{}
\Accepted{}

\title{
  Suzaku and XMM-Newton Observations of Diffuse X-ray Emission 
  from the Eastern Tip Region of the Carina Nebula
}

\author{
  Yuichiro \textsc{Ezoe}\altaffilmark{1},
  Kenji \textsc{Hamaguchi}\altaffilmark{2,3},
  Robert A. \textsc{Gruendl}\altaffilmark{4},
  You-Hua \textsc{Chu}\altaffilmark{4},\\
  Robert \textsc{Petre}\altaffilmark{5}, and
  Michael F. \textsc{Corcoran}\altaffilmark{2,3}
}

\altaffiltext{1}{
  Tokyo Metropolitan University, 1-1, Minami-Osawa, Hachioji,
  Tokyo, 192-0397, JAPAN
}


\altaffiltext{2}{
  CRESST and X-ray Astrophysics Laboratory NASA/GSFC, Greenbelt, MD 20771, USA
}

\altaffiltext{3}{
  Universities Space Research Association,
  10211 Wincopin Circle, Suite 500, Columbia, MD 21044-3432, USA
}

\altaffiltext{4}{
  Department of Astrophysics, University of Illinois, Urbana, IL 61801, USA
}

\altaffiltext{5}{
Astrophysics Science Division, NASA Goddard Space Flight Center, Greenbelt, MD 20771
}


\email{ezoe@phys.metro-u.ac.jp}


\maketitle

\begin{abstract}
The eastern tip region of the Carina Nebula was observed with the Suzaku XIS 
for 77 ks
to conduct a high-precision spectral study of extended X-ray emission.
XMM-Newton EPIC data of this region were also utilized 
to detect point sources.
The XIS detected strong extended X-ray emission from the entire field-of-view 
with a 0.2--5 keV flux of $0.7\sim4\times10^{-14}$ erg s$^{-1}$ arcmin$^{-2}$.
The emission has a blob-like structure that coincides with an ionized gas filament observed
in mid-infrared images.
Contributions of astrophysical backgrounds and 
the detected point sources were insignificant.
Thus the emission is diffuse in nature. 
The X-ray spectrum of the diffuse emission was represented by a two-temperature
plasma model with temperatures of 0.3 and 0.6 keV and an absorption column 
density of 2$\times10^{21}$ cm$^{-1}$. 
The X-ray emission showed normal nitrogen-to-oxygen abundance ratios 
and a high iron-to-oxygen abundance ratio.
The spectrally deduced parameters, such as temperatures and column densities, 
are common to the diffuse X-ray emission near $\eta$ Car. 
Thus, the diffuse X-ray emission in these two fields may have the same origin.
The spectral fitting results are discussed to constrain the origin in the context 
of stellar winds and supernovae.
\end{abstract}

\section{Introduction}
\label{sec:intro}

Diffuse X-ray emission extending over several to tens pc 
has been reported in many massive star-forming regions, such as
NGC 2024 \citep{Ezoe2006a}, 
the Orion Nebula \citep{Guedel2008},
the Rosette Nebula, \citep{Townsley2003},
M17 \citep{Townsley2003,Hyodo2008},
RCW 38 \citep{Wolk2002},
NGC 6334 \citep{Ezoe2006b},
the Carina Nebula \citep{Hamaguchi2007},
W49 \citep{Tsujimoto2006},
NGC 3603 \citep{Moffat2002},
the Arches Cluster \citep{Yusef-Zadeh2002,Tsujimoto2007},
and the Quintuplet Cluster \citep{Wang2002}.
This diffuse component contributes a considerable fraction of the 
total X-ray emission and shows different spectral characteristics 
among these regions.
Diffuse X-ray emission can be roughly classified into three types:
thin-thermal plasma emission with a temperature $kT\sim$ 0.1--1 keV, 
higher-temperature plasma emission with $kT\sim$ 2--10 keV,
and possibly non-thermal emission with a photon index of 1-1.5.
These phenomena can be explained by plasma heating and particle 
acceleration in strong shocks by fast stellar winds from young 
OB stars \citep{Townsley2003,Ezoe2006a,Ezoe2006b,Guedel2008}
and/or past supernova remnants (SNRs) \citep{Wolk2002,Hamaguchi2007}.
The precise origin of diffuse X-ray emission, however, is often
unclear.

Recently \citet{Hamaguchi2007} suggest that the origin of diffuse 
X-ray emission can be constrained by plasma diagnostics or measurements
of elemental abundances.
While main-sequence late-O to early B stars have nearly solar abundances
(e.g., \cite{Cunha1994,Daflon2004}), evolved stars show non-solar elemental 
compositions due to the CNO cycle.
For instance, the plasma will be overabundant in nitrogen if its origin 
is the wind from a nitrogen-rich Wolf-Rayet 
star. 
On the other hand, the plasma will be overabundant in oxygen, neon, and 
silicon if it is produced by a Type II SNR (e.g., \cite{Tsujimoto1995}).
The low-temperature ($kT\sim$0.1--1 keV) type of diffuse X-ray 
emission is ideal for such diagnostic studies, because a variety 
of K-shell lines exist in the 0.2--2 keV range.

The Carina Nebula is an excellent site to investigate 
plasma diagnostics of diffuse X-ray emission.
At a distance of 2.3 kpc, it is one 
of the most active massive star forming regions in the Galaxy.
It contains eight massive stellar clusters: Trumpler 14, 15, 16, 
Collinder 228, Bochum 10, 11, NGC 3293, and NGC 3324.
In total, there exist more than 64 O stars \citep{Feinstein1995,
Smith2006}, including the extreme-type luminous blue variable
$\eta$ Car and four Wolf-Rayet stars.
%
%
%
%
The age of the nebula is estimated to be $\sim3$ Myr based 
on the most evolved stars and the size of the HII region 
\citep{Smith2000}.
The young massive stars that are still enshrouded in gas and dust
have been observed in optical, infrared, and radio wavelengths 
(e.g., \cite{Harvey1979, Smith2000, Yonekura2005}).
The number counts of the most massive O stars, e.g., O3 stars,
suggest that the star-formation activity of the Carina Nebula 
rivals those of the most active regions, such as NGC 3603 
at $D=6.9$ kpc and W49 at $D=11.4$ kpc.
The proximity of the Carina Nebula, compared with NGC 3603 and W49,
makes it the best target to study diffuse X-ray emission
resulting from star-forming activities.

Seward et al.\ (1979) first suggested the existence of possible
diffuse soft X-ray emission in the Carina Nebula with a luminosity of
$\sim$10$^{35}$ erg s$^{-1}$ and a spatial extent of several pc,  
using {Einstein} observations.
Although the limited energy and spatial resolution of Einstein 
hindered the determination of precise plasma parameters and 
the contribution from point sources, \citet{Seward1982}
postulated that the extended X-rays were from hot gas with 
$T\sim10^{7}$ K. 
With {Chandra}, Evans et al.\ (2003) confirmed the existence
of diffuse X-ray emission near $\eta$ Car in addition to 
point sources; however, the limited photon statistics and 
high background prevented detailed spectral analysis.
Hamaguchi et al.\ (2007) obtained spectra
of the diffuse X-ray emission around $\eta$ Car with {Suzaku}.
Owing to the good low-energy response and the low background 
of the X-ray CCD onboard {Suzaku}, 
the spectra extracted from the regions north and south of 
$\eta$ Car can be modeled with high precision; both spectra
are best represented by three-temperature plasma models with 
$kT\sim0.2$, $\sim0.6$, and $\sim5$ keV.
Analyzing the {Suzaku} data in conjunction with XMM-Newton and 
Chandra data, Hamaguchi et al.\ (2007) concluded that the 0.2 keV
and 0.6 keV components most likely originated from diffuse plasma, 
but the 5 keV component could not be easily distinguished from
the unresolved point sources.
They found that the iron and silicon abundances were significantly 
different in the north and south regions, and that the nitrogen-to-oxygen 
abundance ratios in both regions were far lower than those of stellar winds 
from evolved massive stars such as $\eta$ Car and WR25 in this field.
From these results, they concluded that the diffuse X-ray emission
near $\eta$ Car originated from one or multiple SNRs.

We have studied extended X-ray emission 
from an eastern tip region of the Carina Nebula.
Located $\sim$ 30$'$ ($\sim$ 20 pc) from $\eta$ Car, this region is 
less contaminated by X-ray emission from OB stars than the regions 
near $\eta$ Car.
Previous {Einstein} observations have revealed strong extended X-ray 
emission in this region, although no massive stars earlier than B3 
are known here \citep{Seward1982}.
It is not known whether this emission is truly diffuse and, 
if so, whether its origin is similar to that of the regions near $\eta$ Car.
In this paper we report the first detailed spectral analysis of 
the extended X-ray emission in this eastern tip region of the 
Carina Nebula using Suzaku observation.  To augment the limited 
angular resolution of Suzaku, the analysis also made use of
{XMM-Newton} data.

\section{Suzaku Observation}
\label{sec:obs}

Suzaku is the 5th Japanese X-ray observatory \citep{Mitsuda2007}.
It carries four scientific instruments; 
X-ray optics or the X-Ray Telescope (XRT: \cite{Serlemitsos2007}); 
an X-ray calorimeter (XRS: \cite{Kelly2007});
four X-ray CCDs (XIS: \cite{Koyama2007});
and a hard X-ray detector (HXD: \cite{Takahashi2007,Kokubun2007}).
The XIS consists of three front-illuminated (FI) CCDs 
(XIS0, 2, and 3) and one back-illuminated (BI) CCD (XIS1).
Due to the low-earth orbit of Suzaku and the large effective 
area, the XIS has the lowest particle backgrounds 
among all X-ray CCDs in currently available  X-ray observatories.
Furthermore, the XIS has good energy resolution
and superior low-energy response with negligible low-energy tails,
compared to the other X-ray CCDs onboard XMM-Newton and Chandra.

We observed the eastern tip region of the Carina Nebula 
with Suzaku on 2006 June 5. 
During the observation, the XIS and HXD were operated in the normal 
mode. In the present paper, we use only the XIS data because we are 
interested in the spectral analysis of the soft extended X-ray emission.

The data reduction was performed on the version 1.2.2.3 
screened data provided by the Suzaku processing facility,
using the HEAsoft analysis package ver 6.1.1.
No background flares were seen in the data.
The net exposure of each FI and BI chip was 77 ks.
For spectral fits, we generated response matrices and auxiliary 
files with {\tt xisrmfgen} and {\tt xissimarfgen} released on 
2006 October 26.
The in-flight gradual degradation of energy resolution and absorption 
due to XIS contamination were considered in these softwares.

\section{Extended X-ray Emission}
\label{sec:diff}

\begin{figure}[p]
  \begin{center}
    \FigureFile(0.7\textwidth,){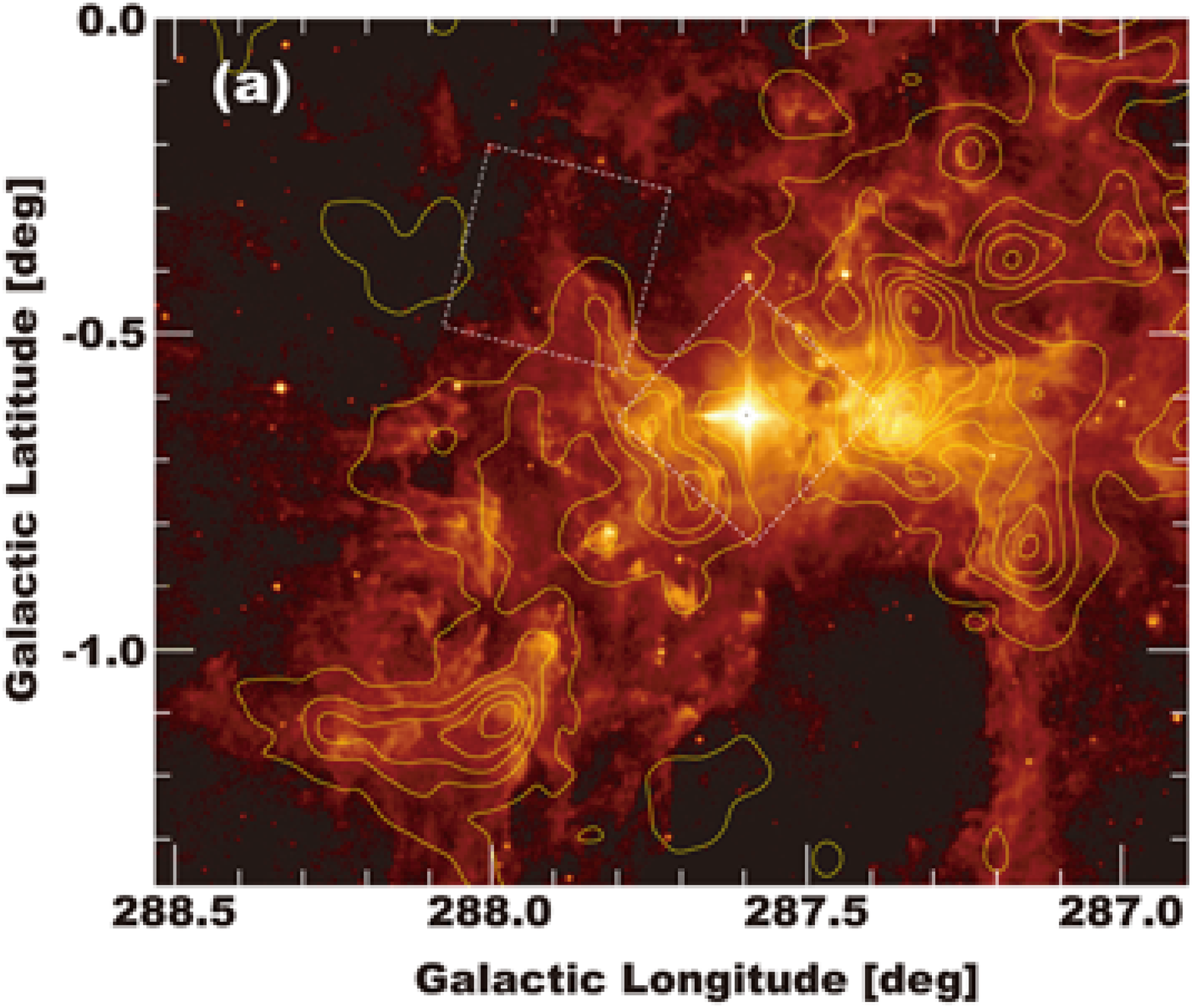}
    \FigureFile(0.7\textwidth,){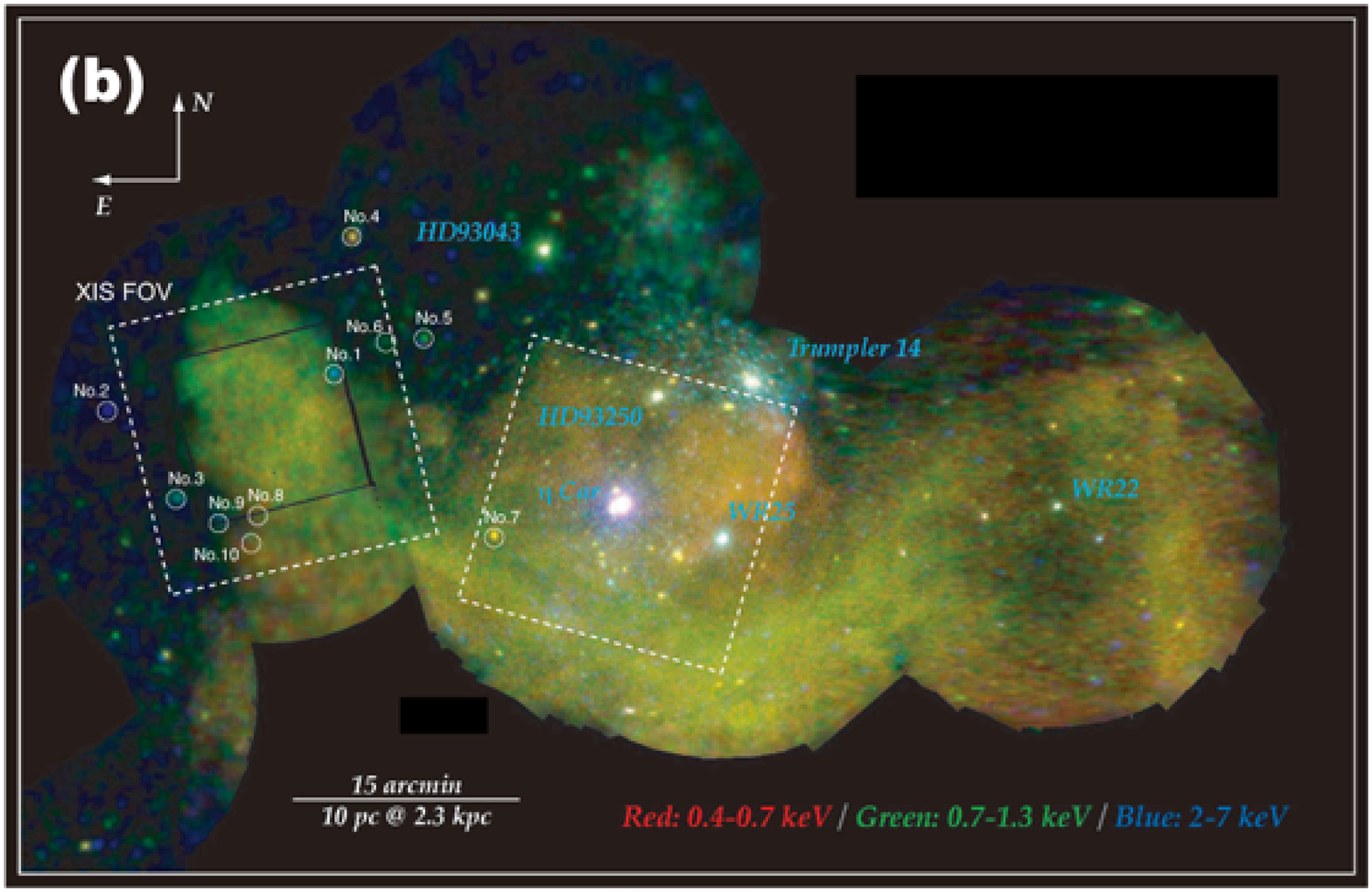}
  \end{center}
  \caption{(a) 
                   An MSX band-A 8.28 $\mu$m image of the Carina nebula
                    taken from the NASA/IPAC Infrared Science Archieve.
                    Yellow contours show the 12CO (J$=$2-1) map 
                    \citep{Yonekura2005}.
                    Two white boxes represent field-of-views of this 
	           (upper left) and previous Suzaku observations (lower right, 
	            \cite{Hamaguchi2007}).
	           (b) An EPIC MOS mosaic of the same nebula 
	            with XMM-Newton taken from the XMM-Newton Image Gallery.
	           The red, green and blue colors show soft (0.4--0.7 keV), 
	           medium (0.7--1.3 keV) and hard (2--7 keV) X-ray emission, 
	           respectively.
	   	  Circles and numbers indicate point sources detected in our 
	           analysis (see \S \ref{sec:ps}). 
  }\label{fig:overview}
\end{figure}

Figure \ref{fig:overview}a shows the location of our observation
(top left box) on an MSX 8.28 $\mu$m image of the Carina Nebula 
retrieved from the NASA/IPAC Infrared Science Archive 
\footnote{http://irsa.ipac.caltech.edu/applications/MSX/MSX/}.
This mid-IR image is dominated by cationic polycyclic aromatic 
hydrocarbon emission in photodissociation regions \citep{Smith2000}
and traces the surface of molecular clouds that are ionized by 
stellar winds and ultra-violet radiations from OB stars. 
A curved mid-infrared filament runs across the XIS field-of-view (FOV).
Figure \ref{fig:overview}b shows an X-ray overview of the same
area with XMM-Newton MOS\footnote{http://xmm.esac.esa.int/external/xmm\_science/gallery/}.
In X-rays, there exists patchy extended soft X-ray emission whose
distribution roughly follows the mid-infrared filament.
This suggests that the X-ray emission contributes to the ionization
of the molecular cloud and that the emitting hot gas is in contact 
with the surface of the molecular clouds.

\begin{figure}[p]
  \begin{center}
    \FigureFile(0.8\textwidth,){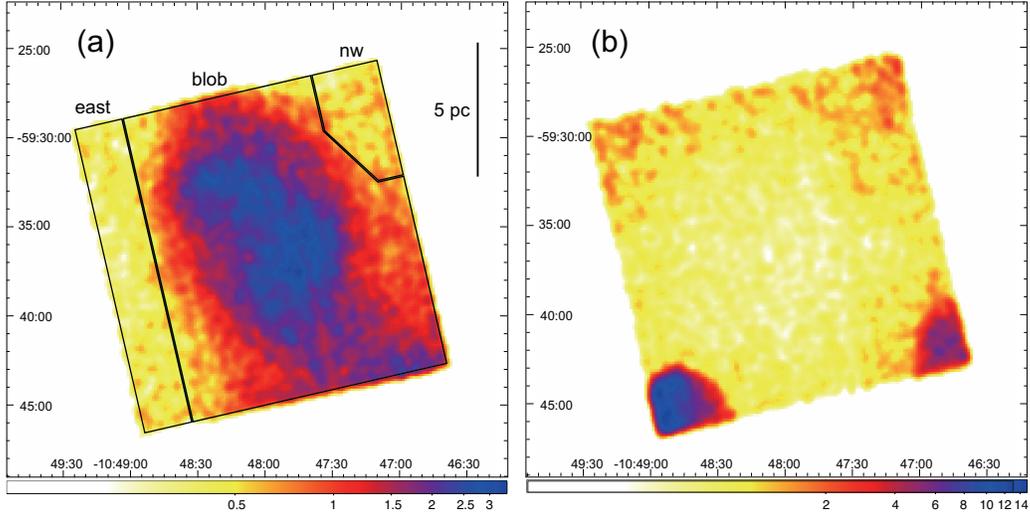}
  \end{center}
  \caption{Suzaku BI images of the eastern tip region of the Carina 
           nebula in the (a) 0.2--2 keV and (b) 2--10 keV bands, 
	   displayed on the J2000.0 coordinates.
	   For clarity, images are binned by a factor of 8 and smoothed 
	   by a Gaussian of $\sigma$ = 3 pixels. 
	   Vignettings are corrected (see \S \ref{sec:obs}).
	   The unit of intensity in the greyscale wedge is arbitrary.
	   Solid black lines mark regions utilized in the 
	   spectral analysis. 
	   The strong emission at the bottom-left and -right parts of
           the panel (b) corresponds to calibration 55Fe sources.
  }\label{fig:xisimage}
\end{figure}

We created Suzaku XIS images in 0.2--2 and 2--10 keV bands, 
as shown in figure \ref{fig:xisimage}.
The vignetting in the images has been corrected by dividing
the observed images by model XIS images produced with the XRT$+$XIS 
simulator {\bf xissim} for a uniform surface brightness.
In simulations, we assumed monochromatic X-rays of energy 
1.49 keV and 8.05 keV, since the vignetting is best studied 
in these energies \citep{Serlemitsos2007}.
A blob-like extended X-ray emission is clearly detected 
in the soft X-ray band.
No significant X-ray emission is seen in the hard X-ray band, which
is consistent with the XMM-Newton image (figure~\ref{fig:overview} b).

\begin{figure}[p]
  \begin{center}
    \FigureFile(0.55\textwidth,){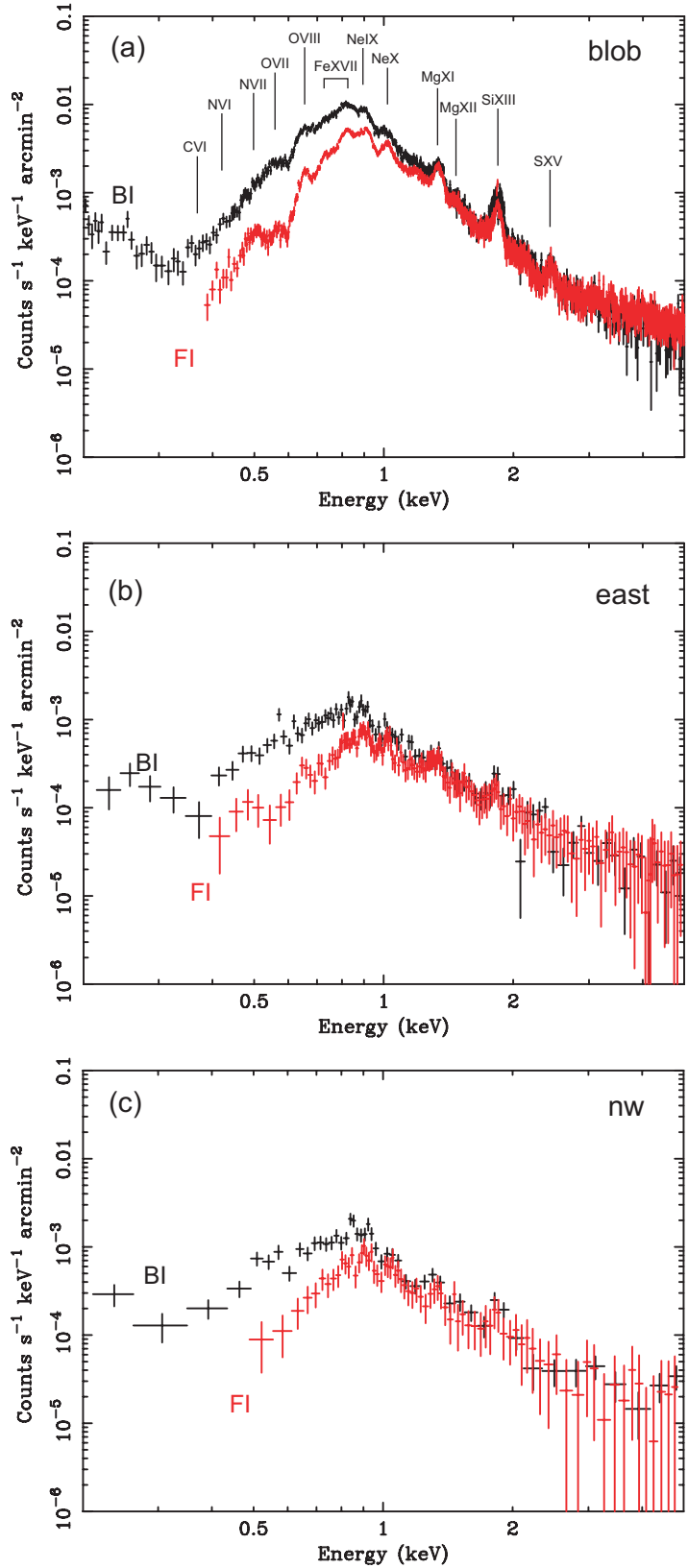}
  \end{center}
  \caption{Background-subtracted BI (black) 
    and FI (red) spectra of (a) the blob, 
    (b) the east, and (c) the nw regions. Center energies of 
    emission lines are shown in the panel (a). 
    For comparison, the vertical axis is normalized by the 
    angular size of each region.
  }\label{fig:xisdiffspec}
\end{figure}

To evaluate the significance of the extended X-ray emission, 
we defined a region named blob, as shown in figure~\ref{fig:xisimage}.
The total area of the blob region is 234 arcmin$^2$, or 105 pc$^2$. 
We extracted the XIS0-3 spectra from this region. 
To evaluate the background, we utilized spectra accumulated 
from observations of the night side of the Earth, as the night
Earth backgrounds reproduced well all the observed spectra above 
$\sim$6 keV, where the instrumental background dominated.
The background-subtracted XIS1 (BI) and XIS$0+2+3$ (FI) 
spectra are shown in figure \ref{fig:xisdiffspec}a, 
and their 0.2--5 keV count rates with 1$\sigma$ uncertainties 
are $1.073\pm0.004$ (BI) and  $0.609\pm0.002$ (FI) counts s$^{-1}$, 
respectively.
The emission is highly significant and shows a 
number of emission lines such as O VII, O VIII, 
Ne IX, Ne X, Mg XI, Si XIII, and S XV.
Lines from these ions in different ionization states,
such as O VII and Si XIII, suggest that the extended X-rays 
are from not a single-temperature but multi-temperature plasma.

We extracted spectra from two surrounding regions named 
east and nw, as shown in figure~\ref{fig:xisimage}. 
The areas of the east and nw regions are 47 arcmin$^2$ (or 21 pc$^2$)
and 21 arcmin$^2$ (or 9.3 pc$^2$), respectively.
We again used the night Earth spectra at the respective detector 
positions as backgrounds. The background-subtracted spectra 
are shown in figures \ref{fig:xisdiffspec}b and c.
Although their surface brightnesses are an order-of-magnitude 
lower than that in the blob region, there exist signs of 
emission lines from O VII, Ne IX, Ne X, Mg XI, and Si XIII.
Thus extended X-ray emission appears to be present over 
the entire FOV.
The surface brightness below 0.3 keV and above 2 keV 
is similar in all three regions.
Therefore, in addition to the plasma emission that is 
dominant between 0.3 and 2 keV, there must be 
additional X-ray sources.
Plausible candidates are the local hot bubble (LHB), 
the cosmic X-ray background (CXB), the galactic ridge 
X-ray emission (GRXE), and point sources.
LHB and CXB are uniform and common 
background sources existing in all X-ray observations.
GRXE is also a uniform X-ray background and must 
contribute to the emission because the observational
FOV is located in the Galactic plane 
($l=288^\circ$, $b=-0^\circ\hspace*{-1mm}.4$,
see figure~\ref{fig:overview}a).
The point sources, on the other hand, are position-dependent.
Contamination from these astrophysical backgrounds
and point sources needs to be carefully considered in order to
characterize the possible diffuse emission from the Carina
Nebula.

\section{Point Sources}
\label{sec:ps}

To quantify the contribution from point sources, 
we analyzed the XMM-Newton data.
XMM-Newton observed this region on 2004 December 7 for 27 ks.
The European Photon Imaging Camera (EPIC) provided CCD imaging
spectroscopy with one pn  camera \citep{Struder2001} and two MOS
cameras \citep{Turner2001}.
The medium optical blocking filter was used. 
As shown in figure \ref{fig:overview}b, the FOVs of MOS cover 
the entire FOV of the XIS.
We analyzed the archival processed data using SAS (Science Analysis 
Software) version 7.0.0, following the SAS user's 
guide\footnote{http://xmm.esac.esa.int/external/xmm\_user\_support/documentation/sas\_usg/USG/}.
The event files were time-filtered to exclude periods of high 
background, during which the count rate from the entire CCD area 
at energies $>$10 keV is more than 1.2 times the average rate for
each MOS and pn observation.
This removed $\sim$1 ks from each observation, and yielded 22 and 
26 ks\sout{ec} of usable exposure for pn and each MOS, respectively.
Source detection was accomplished with the SAS program 
{\tt edetect\_chain}.
Images in 8 bands (0.2--0.5, 0.5--2.0, 2.0--4.5,
4.5--7.5, 7.5--12.0, and 0.2--12.0 keV) were utilized, 
in order not to miss very soft or very hard sources.
The resulting source list was checked manually for spurious 
detections and missed sources.
As a result, 10 sources have been detected, among which
5 sources were within the FOV of the XIS.
The locations of these X-ray point sources are marked by
circles in figure \ref{fig:overview}b.

For each individual point source, we extracted counts using a circular 
region of a 30 arcsec radius centered at the source.
Background counts were extracted from an adjacent annular region 
with an outer radius of 1 arcmin.
In the case that the annular region included other sources, the 
background was then extracted from a nearby circular source-free region.
To identify counterparts of these X-ray point sources in other 
wavelengths, we searched the 2$\mu$m all sky survey catalog (2MASS)\footnote{http://irsa.ipac.caltech.edu/index.html}
and the AXAF Guide and Acquisition Star Catalog (AGAST)\footnote{http://cxc.harvard.edu/cgi-gen/cda/agasc/agascInterface.pl}
for candidates within 10 arcsec, the angular resolution of XMM-Newton.
In cases where multiple candidates exist, we chose the closest one as the
most plausible counterpart.
Among the 10 X-ray sources, 9 have counterparts in 2MASS 
and 1 has a counterpart in AGAST.
The properties of these individual sources are summarized in table \ref{tbl:srclist}.

We examined the 10 sources for possible temporal variations. 
For each source, we produced pn and MOS X-ray light curves
in the 0.4--10 keV band using a binning size of 512 s bin$^{-1}$. 
The light curves were examined against a constant hypothesis 
in terms of $\chi^2$ statistics. 
If the $\chi^{2}$ probability of constancy became less than 4\%, 
at least in one detector, we regarded the source as variable.
Only the source No.3 has been found to be variable.
It showed a factor of about 5 increase in the first 5 ks of the 
observation, decreased in the next 5 ks, and stayed constant in the rest.
Such rapid temporal variations strongly suggest that No.3 
is a young low-mass star.

\begin{figure}[p]
  \begin{center}
    \FigureFile(0.8\textwidth,){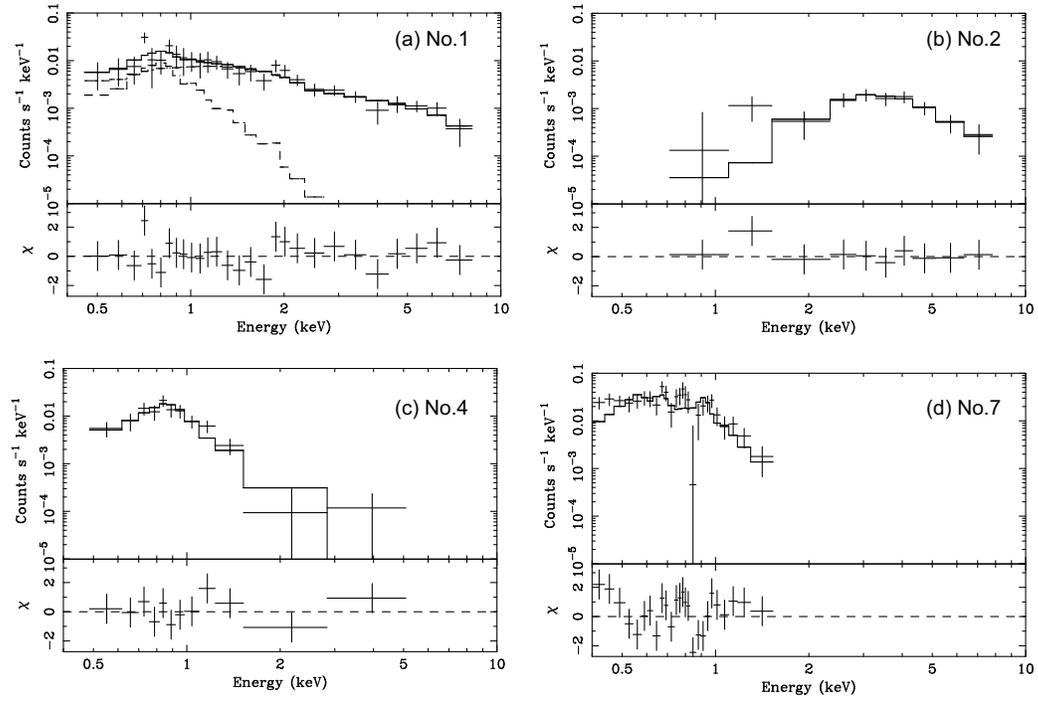}
  \end{center}
  \caption{XMM-Newton EPIC-pn spectra of the bright point sources
    (No.1, 2, 4, and 7).
    The solid lines show the best-fit absorbed plasma models. 
    The dotted lines in the panel (a) show two plasma components. 
    The bottom panels exhibit residuals from the best-fit models.
  }\label{fig:xmmsrcspec}
\end{figure}

We also conducted spectral analysis for 4 bright sources 
(No.1, 2, 4, and 7) that had $>100$ counts in the pn observation.
We analyzed only the pn spectra because of the limited statistics 
of MOS. 
The SAS tasks {\tt rmfgen} and {\tt arfgen} were utilized to 
generate response matrix files and auxiliary files for each source.
We fitted the spectra using a thin-thermal plasma emission model 
in collisional equilibrium (the APEC mode; \cite{Smith2001}) 
convolved with the interstellar absorption.
Such models, with abundances fixed at 0.3 solar, are commonly 
used in X-ray spectral analyses of star-forming regions 
(e.g., \cite{Getman2005}).
We found that all the spectra except that of No.1 were well represented 
by this simple model.
For the No.1 spectrum, we tried a two-temperature plasma emission model 
with a common absorption, and were able to find acceptable fits.
The fitting results are summarized in table \ref{tbl:srcfit} and 
figure \ref{fig:xmmsrcspec}.

Source No.1 has a very hard continuum without any sign of emission 
lines, thus it may be a background active galactic nucleus.
Source No.2 may be an embedded young low-mass star because 
it has a large absorption column density and a moderate temperature,
while source No.4 may be a foreground star because of its
small absorption column and moderate temperature.
Source No.7 is peculiar with a low temperature and possibly high luminosity.
Its position is within 8 arcsec from the X-ray source No.78 in an
XMM-Newton observation of $\eta$ Car reported by \citet{Colombo2003}.
Further optical spectroscopic study is needed to identify the nature 
of this source.

We estimated X-ray fluxes of the other 6 sources assuming a 
thin-thermal plasma model and convolving it with an estimated interstellar 
absorption of $N_{\rm H}=1.3\times10^{22}$ cm$^{-2}$.
We adopted a temperature of 3 keV and a metal abundance of 0.3 solar
that are typical of emission from young stars (e.g., \cite{Imanishi2001}).
WebPIMMS\footnote{http://heasarc.gsfc.nasa.gov/Tools/w3pimms.html}
was used to convert the pn count rates to X-ray fluxes. The results are shown 
in table \ref{tbl:srclist}.
The fluxes range from 2$\times$10$^{-14}$ to 4$\times$10$^{-13}$
erg s$^{-1}$ cm$^{-2}$.

\section{Contamination from Astrophysical Backgrounds
and Point Sources}
\label{sec:contami}

We proceeded to estimate contribution 
to the observed X-ray emission by 
LHB, CXB, GRXE, and point sources.
For the LHB, CXB, and GRXE,
we assumed the same models in \citet{Hamaguchi2007}:
the Raymond-Smith thin-thermal plasma model with 
$kT\sim0.1$ keV and a surface brightness of $\sim4\times10^{4}$
counts s$^{-1}$ arcmin$^{-2}$ for LHB based on \citet{Snowden1998},
the model Id1 in table 2 of \citet{Miyaji1998} for CXB, and
the free abundance model in table 8 of \citet{Ebisawa2005}
with the X-ray flux of $1.4\times10^{-11}$ erg cm$^{-2}$ 
s$^{-1}$ deg$^{-2}$ (3--20 keV) for GRXE.
We used the same spectral parameters except 
for the normalization to fit, which was adjusted.
To estimate the X-ray fluxes from these components with
different spatial distribution, 
we prepared an arf file for
the uniform emission (LHB, CXB, and GRXE)
by using the {\tt xissimarfgen} program, while
we created an arf file for each point source.
We used the best-fit models in table \ref{tbl:srcfit}
for the bright point sources and assumed the typical
spectral model of young low-mass stars for the other
faint point sources ($kT=3$ keV and $N_{\rm H}=1.3\times10^{22}$ cm$^{-2}$)
using the APEC emission code.

\begin{figure}[p]
  \begin{center}
    \FigureFile(0.55\textwidth,){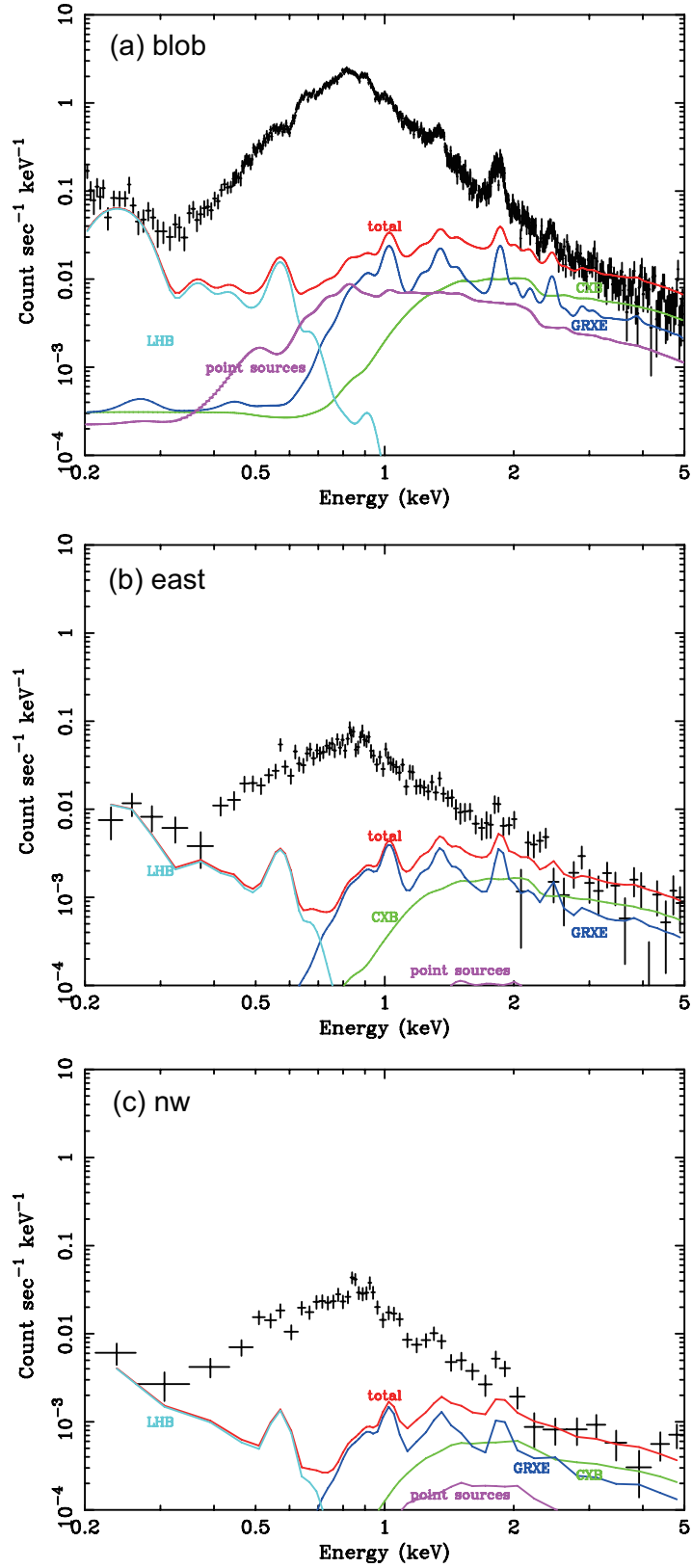}
  \end{center}
  \caption{Background-subtracted BI spectra of 
    (a) the blob, (b) the east, and (c) the nw regions.
    Solid lines show estimated contamination from 
    X-ray sources.
  }\label{fig:bgdest}
\end{figure}

In figure \ref{fig:bgdest}a, we plot the 
estimated contamination from LHB, CXB, GRXE, and 
point sources in the blob region.
Below 0.3 keV, LHB accounts for the X-ray emission, while
X-rays above 2 keV can be explained by the sum of CXB, 
GRXE and point sources.
This makes a sharp contrast to the $\eta$ Car region 
where a residual hard X-ray emission is seen above 2 keV.
The excellent fit of these sources to the spectrum 
below 0.3 keV and above 2 keV supports the validity 
of our estimation.
Thus, almost all the excess counts in 0.3--2 keV can be 
considered to be truly diffuse plasma emission.
This conclusion is also supported by the clear spectral 
differences between the diffuse plasma and point sources.

In the same way, we estimated the X-ray contamination 
in the east and nw regions as shown in figures 
\ref{fig:bgdest}b and c.
In both regions, LHB, CXB, GRXE, and point sources 
explain all the emission in $<0.3$ and $>2$ keV 
as well as in the blob region, but there are still 
excesses in 0.3--2 keV.
Therefore, diffuse soft X-ray emission exists in these fields, too.

\section{Spectral Analysis}
\label{sec:spec}

\subsection{The blob Region}
\label{sec:spec:blob}

The nature of the diffuse X-ray emission is investigated
through spectral analysis.
We simultaneously fitted the 0.2--5 keV XIS 
BI and FI spectra of the blob region.
We created XIS arf files using {\tt xissimarfgen},
assuming the 0.4--2 keV XMM MOS image in figure \ref{fig:overview}b
as spatial distribution of the diffuse X-ray emission.
In single or two-temperature plasma models utilized below, 
we allow the abundances of the noticeable elements (O, Ne, Mg, Si, 
S, and Fe) to vary, while those of the other elements were 
fixed at 0.3 solar value, which is generally seen in low-resolution
CCD spectra of young stars.

In addition to the plasma model for the diffuse X-ray emission, 
we introduced a thin-thermal plasma model and a power-law 
model to reproduce the LHB, CXB, GRXE, and point source
contributions (\S \ref{sec:contami}).
For the plasma model of LHB, we fixed the temperature
at 0.1 keV and the abundances at 1 solar, but allowed
the normalization to vary.
For simplicity, the CXB, GRXE, and point sources were together
approximated by a single power-law model with the photon
index fixed at 1.5 and convolved with the same absorption 
of the diffuse X-ray emission.
To take into account possible uncertainties in the energy 
scale calibration, we introduced two additional fitting
parameters, gain and offset.
Throughout the fittings below, the best-fit energy scale 
and offset values were $<$1 \% and $<$6 eV, respectively, 
consistent with the current calibration uncertainties
\footnote{http://www.astro.isas.jaxa.jp/suzaku/process/caveats/}.

\begin{figure}[p]
  \begin{center}
    \FigureFile(0.55\textwidth,){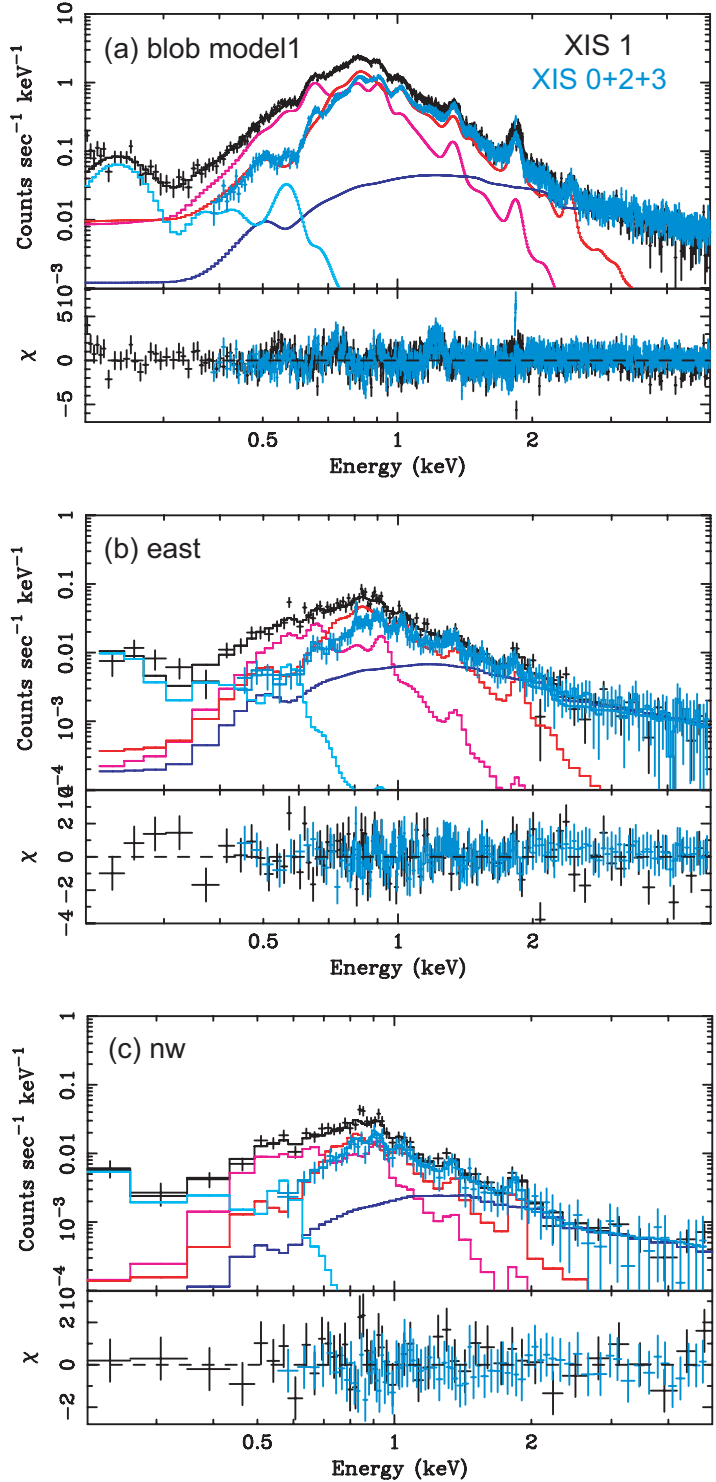}
  \end{center}
  \caption{Best-fit spectral results of the diffuse X-ray emission
      in (a) the blob, (b) the east, and (c) the nw regions. 
      The best-fit model is shown in a solid black line. The model 
      components for the XIS1 spectrum are shown in solid colored lines 
      (cyan: the LHB component, magenta and red: the two-temperature 
      plasma component, blue: the power-law component).
      See tables \ref{tbl:fit1} and \ref{tbl:fit2} 
      for the obtained parameters.
  }\label{fig:fit12}
\end{figure}

We first tested a single-temperature thin-thermal 
plasma model convolved with an absorption.
This simple model yielded an unacceptable best fit 
for the spectra with $\chi^2$/d.o.f. of $2.4$.
The best-fit temperature, $kT$ = 0.59 keV, was too 
high to reproduce the significant OVII and NeIX lines.
We then tried a commonly-absorbed 
two-temperature plasma model as shown in figure 
\ref{fig:fit12}a and table \ref{tbl:fit1} (model 1).
This model represents the data far better ($\chi^2$/d.o.f. of $1.2$). 
A small discrepancy of the fit to the data at 0.5--0.8 keV
and 1.1--1.2 keV may be caused by inaccuracy of the Fe 
L-shell emission line model and calibration uncertainty 
near the mirror Au L edges, respectively.   
Indeed, similar discrepancies at 0.5--0.8 keV can be seen in
other Suzaku XIS spectral fits (e.g., \cite{Hamaguchi2007}).
Therefore, we think the best-fit two-temperature model 
represents the data well, although the $\chi^2$ is still 
not above the 90\% confidence level.

\subsection{The east and nw Regions}
\label{sec:spec:eastnw}

Similar to the analysis of the blob region, we fitted the east 
and nw spectra with a two-temperature plasma model.
Since no evident spatial structures were seen in these regions 
(figure \ref{fig:xisimage}),
we created arf files assuming a uniform emission.
We did not use the additional fitting parameters gain and offset
because  the photon statistics were limited.
The results are shown in figures \ref{fig:fit12}b and c, and table \ref{tbl:fit1}. 
In both regions, the fits are acceptable and all resulting 
parameters except the surface brightness are consistent with 
those in the blob region, although the uncertainties are large.
As expected from figure \ref{fig:xisimage}, the surface brightness 
of the east and nw regions is a factor of $\sim$5 lower than 
that in the blob region.

We also tried a single-temperature plasma model and 
obtained acceptable fits in both regions, but the best-fit 
column densities were small, $\sim8\times10^{20}$ cm$^{-2}$,
and the best-fit temperatures were high, $\sim$0.6 keV.
Such a large variation in the column density within the XIS FOV
is inconsistent with the CO map (figure \ref{fig:overview}a), 
although the CO gas can lie behind the X-ray emitting gas.
There is also a hint of OVII K emission in both spectra 
(figures \ref{fig:xisdiffspec} b and c) that cannot be reproduced
by this best-fit single-temperature model.
Therefore, it is likely that the east and nw spectra are also
best fitted by a two-temperature plasma model.

\subsection{XMM-Newton spectra}
\label{sec:spec:xmm}

\begin{figure}[p]
  \begin{center}
    \FigureFile(0.55\textwidth,){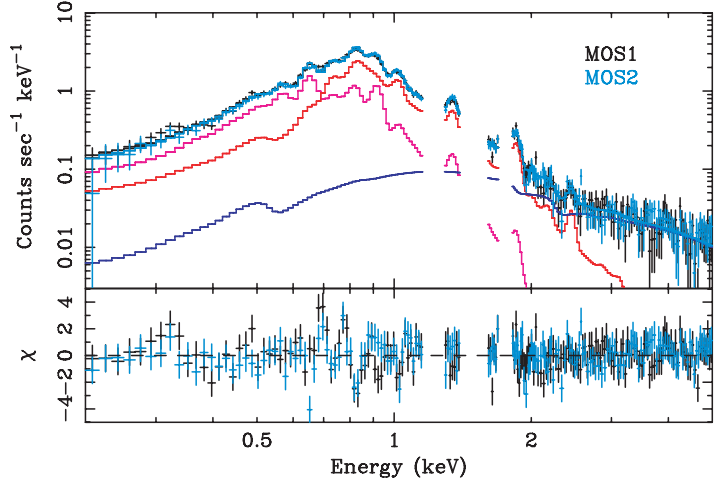}
  \end{center}
  \caption{XMM-Newton MOS fitting result of the diffuse X-ray emission
      in the blob region. 
      Line colors are the same as figure \ref{fig:fit12} but for 
      the MOS1 and MOS2 data.
      See table \ref{tbl:fit3} for the obtained parameters.
  }\label{fig:fit3}
\end{figure}

We also analyzed the XMM-Newton observations of the blob region
with the same data used in the point source analysis (\S \ref{sec:ps}).
We used only the MOS data with relatively low particle background events,
whose background spectrum during the observation can be easily estimated
from the ESAS package\footnote{http://heasarc.gsfc.nasa.gov/docs/xmm/xmmhp\_xmmesas.html}.
We used SAS version 7.0.0 and ESAS version 1.0 for the analysis, and
generated response matrices using the {\tt rmfgen} and {\tt arfgen}.
We found that the two-temperature plasma model well represents 
the MOS spectra, as was the case for the Suzaku XIS spectra.
Figure \ref{fig:fit3} and table \ref{tbl:fit3} show the fitting result. 
Since there can be C K$_\alpha$ emission around 0.4 keV,
we allowed the C abundance to vary.
The energy band of 1.16--1.28, 1.4--1.6, and 1.7--1.8 keV 
are omitted in order to exclude the instrumental emission lines
from Mg K$_\alpha$, Al K$_\alpha$, and Si K$_\alpha$.
Because the low energy tail of the MOS response 
prevents the distinction between the diffuse 
low temperature plasma component and the LHB, 
only an upper limit is obtained for the LHB. 
Alternatively the best-fit column density becomes somewhat 
lower than that in table \ref{tbl:fit1} (model 1),
to compensate the decreased low energy counts by the LHB component.
The other two-temperature plasma parameters such as 
temperatures and abundances are surprisingly similar
to those in table \ref{tbl:fit1} (model 1).
The only difference is the higher power-law component  flux.
This may be caused by the relatively higher background of 
XMM-Newton observations and their higher uncertainty.
We thus conclude that the XIS and MOS data are consistent 
with each other and the fitting result of the XIS data is reliable.

\section{Discussion}
\label{sec:diss}

We investigated extended X-ray emission in the eastern tip region of the 
Carina Nebula with Suzaku XIS. 
For the first time, we conducted detailed spectral analysis of the 
X-ray emission and found that there is indeed diffuse X-ray emission,
even considering LHB, CXB, GRXE, and point sources.
The diffuse X-ray emission is well represented by a two-temperature 
plasma model with $kT$ $\sim$0.25 and $\sim$0.55 keV. 
Emission measures and abundances of O, Ne, Mg, Si, S, and Fe 
are well constrained owing to the good photon statistics and 
the excellent energy response of XIS.
Below we estimate plasma properties based on the 
spectral fitting and then compare the spectral parameters 
such as absorption column density, temperature and abundance
to those in the $\eta$ Car region, in order to assess the 
origin of the diffuse plasma.

\subsection{Physical Properties of the Plasmas}
\label{sec:diss:prop}

High signal-to-noise XIS spectra enable us to accurately 
constrain the parameters of the diffuse plasma in the eastern 
tip region of the Carina Nebula.
The best-fit column density of $N_{\rm H}\sim2$--$3\times10^{21}$ cm$^{-2}$
and also the temperatures of $kT\sim0.2$--0.3 keV and 0.4--0.6 keV 
in the blob, east and nw regions agree very well with those in the $\eta$ 
Car regions (see the medium-temperature components of the $N_{\rm H}$, $kT$ 
tied model in table 2 of \cite{Hamaguchi2007} or hereafter H2007 model; 
$N_{\rm H}\sim2\times10^{21}$ cm$^{-2}$, $kT\sim0.2$ keV and $0.6$ keV).
Although the high-temperature component represented by a hot plasma model
with $kT\sim5$ keV in the H2007 model is not seen in the eastern
tip region, this is thought to be a composite of diffuse hard X-ray 
emission and contaminations from CXB, GRXE, and point source, and hence
can be ignored.  
This similarity in basic plasma parameters provides strong evidence that 
the diffuse plasma in the vicinity of the Carina Nebula has the same origin.

The continuous distribution of the diffuse X-ray emission over the 
Carina Nebula, as shown in figure \ref{fig:overview}, supports this hypothesis. 
Thus, we can draw a scenario that the X-ray emitting diffuse plasma generated 
by stellar winds from OB stars and/or SNRs forms hot bubbles with the size of 
several tens of pc and ionizes ambient molecular clouds which can be seen in 
the mid-infrared emission.

To investigate the origin of the diffuse plasma,
we estimate physical properties of the plasma in the blob region. 
We assume that the plasma in the blob region have a prolate ellipsoidal 
shape with the major and minor axis lengths of 8 and 3 pc, respectively.
Following the plasma analysis by \citet{Townsley2003}, we estimate the 
electron density, the pressure, the total energy content, the cooling 
time and the mass of the plasma from the observed X-ray luminosity, 
temperatures, and the assumed volume. 
We derived two sets of plasma parameters for the two-temperature 
components from model 1 in table \ref{tbl:fit1}.
The derived parameters are summarized in table \ref{tbl:plasma}.
Below we examine the two interpretations, i.e., stellar winds
from OB stars and SNRs, based on these derived parameters.
Because there are no massive OB stars in the eastern tip region, 
we must consider the massive stellar clusters in the central part 
of the Carina Nebula for the former scenario.

The estimated plasma pressure $P$ is on the order of $\sim$10$^{6}$ 
K cm$^{-3}$  and should be larger than that of the surrounding gas 
in the eastern tip region since both the CO and radio continuum 
intensities are weak in the vicinity of the blob region 
\citep{Yonekura2005, Huchtmeier1975}.
This means that diffuse plasma in the eastern tip region
could flow from its neighbor.
If we consider that the plasma originated in stellar winds from OB stars
near $\eta$ Car and has been 
propagated to this region at the plasma sound velocity, the crossing 
time will be 0.1 Myr. 
Because this time scale is much shorter than that of the radiative 
cooling timescale, $t_{\rm cool}$ $>$1 Myr, the plasma temperatures
can be held constant.
OB stars are able to continuously produce the plasma during 0.1 Myr 
since their typical life time is at least ten times longer.
The SNR interpretation is also possible in terms of the pressure
if one or multiple SNRs occurred in these regions.

The total thermal energy of the plasma, $U$ = $1\times10^{48}$ ergs,
is marginally lower than the total kinetic energy supplied by the 
stellar wind from a single massive star within $\sim1$ Myr, 
$\sim3\times10^{48}$ ergs \citep{Ezoe2006a}.
As the Carina Nebula contains $>$64 OB stars, the observed thermal
energy can be easily supplied by the mechanical energy of the
stellar winds from the $>$64 OB stars, $>2\times10^{50}$ ergs.
If all the diffuse X-ray emission observed with {Einstein} has 
the same origin, the total plasma energy will be about 10 times
larger, i.e., $1\times10^{49}$ ergs. 
Assuming that the stellar winds from these $>64$ OB stars 
($>2\times10^{50}$ ergs) are responsible for the hot plasma, 
the kinetic-to-thermal energy conversion efficiency will be $<$5\%.
According to \citet{Weaver1977}, the thermal energy in the 
shocked stellar wind is 5/11 of the total stellar wind 
kinetic energy. Thus, this conversion efficiency may be
doubled, $<$10 \%, which is comparable to that in M17 ($\sim$10\%, 
\cite{Townsley2003}) 
and larger than that in the Orion nebula ($\sim0.01$\%, 
\cite{Guedel2008}). 
In the SNR case, we can also explain the energy by 
only one canonical supernova ($\sim1\times10^{51}$ ergs) even if 
we must explain all the diffuse X-ray emission in the entire 
Carina Nebula.

The mass of the plasma, $M_{\rm plasma}$ $\sim0.4M_{\odot}$, 
needs at least four typical OB stars assuming a typical mass 
loss rate of stellar winds of 10$^{-7}M_{\odot}$yr$^{-1}$ in 1 Myr. 
If we consider the whole Carina Nebula,  
about $>$80 OB stars are necessary for the entire diffuse 
plasma mass. The known number of OB stars is thus marginal.
On the other hand, one SNR again can supply this mass 
(e.g., \cite{Willingale2003}).

Therefore, although both the stellar-wind and SNR interpretations 
are possible, SNR(s) can explain the derived plasma parameters
such as the total plasma energy and the plasma mass, more easily.
As suggested by \citet{Hamaguchi2007}, the existence of the Carina 
flare supershell \citep{Fukui1999} validates the SNR scenario.
If true, the observed plasma temperatures of 0.3 and 0.6 keV 
limit the SNR age to less than $\sim$10$^4$ yr, since an older SNR 
would be in the radiative phase and efficiently cool down to 
less than 0.1 keV.


\subsection{Abundance}
\label{sec:diss:abd}

\begin{figure}[p]
\hspace*{-10mm}
    \FigureFile(1.1\textwidth,){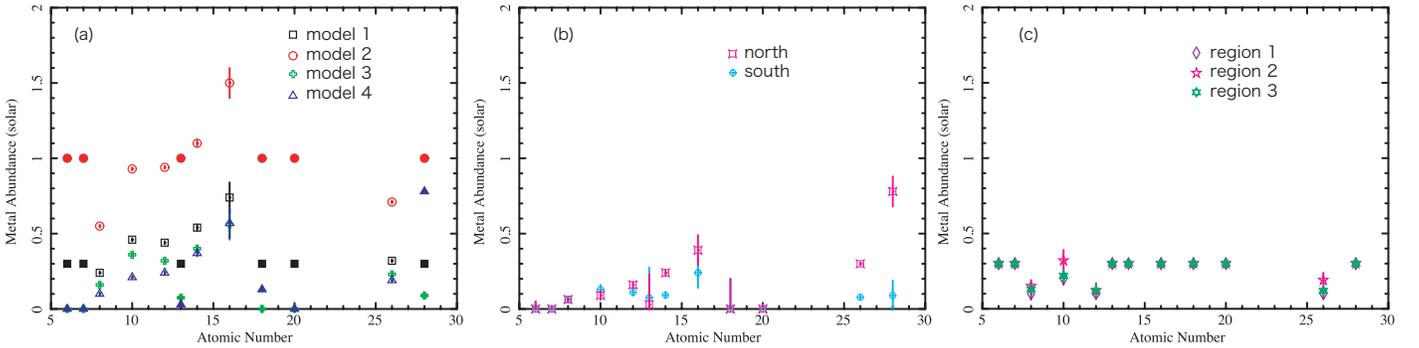}
  \bigskip
  \caption{Abundance distributions in (a) the blob region in the
       eastern tip region (this paper), 
       (b) the north and south regions in the 
       $\eta$ Car region \citep{Hamaguchi2007}, 
       and (c) sub-regions in M17 \citep{Hyodo2008}.
       The filled marks represent the fixed values.
       }\label{fig:abd}
\end{figure}

The abundance pattern of the X-ray emitting plasma provides 
another key piece of information to constrain its origin.
We showed abundance patterns of the blob region and that
of the $\eta$ Car regions in figures \ref{fig:abd}~a and b, 
respectively.
Since \citet{Hamaguchi2007} divided the $\eta$ Car region 
into the north and south regions and fitted the two spectra 
simultaneously with tied column density and temperatures
in the H2007 model, two sets of abundances are shown.

All the metal abundances in the blob region (model 1, black) 
are significantly higher than those in the $\eta$ Car regions.
This may strengthen the result of \citet{Hamaguchi2007}
that the metal abundances of the diffuse X-ray emission 
show spatial variations.
However, there is a possibility that the fixed abundances 
of C, N, Al, Ar, Ca and Ni in table \ref{tbl:fit1} 
influence the other metal abundances.
Thus, we refitted the blob region spectra with different 
fixed abundance sets.

Firstly we tested one solar for the fixed abundances.
In model 1, we implicitly assumed 0.3 solar for them 
because the value is generally used for young stars in star-forming regions.
However, since any mixing of the plasma generated by either
stellar winds or SNR(s) interacting with the ambient molecular clouds 
contains processed stellar material and may have higher abundances,
it is thus worth considering the solar abundances in modeling the 
diffuse emission.
The results are summarized in figure \ref{fig:abd}~a and 
table \ref{tbl:fit1} (model 2).
The $\chi^2$ was comparable to that of model 1 and all 
the abundances increased by a factor of $\sim$2, while 
the emission measures of the 
two plasma components decreased by about the same amount,
to balance the increased line intensities.
Since the SNRs with an age around 10$^4$ yrs, like the Cygnus 
Loop or Vela, globally show no strong deviation from 
solar abundances \citep{McEntaffer2008},
this fitting result allows the SNR interpretation.

Next we assumed the abundances of the H2007 model 
in which all the abundances were allowed to vary and 
constrained with good photon statistics. We used two sets of fixed 
abundances corresponding to the north and the south fits. 
The results are summarized in figure \ref{fig:abd}~a and 
table \ref{tbl:fit1} (models 3 and 4).
In both cases, the fittings were acceptable and all the 
free abundances were significantly decreased.
These changes were caused by the decreased N and Ni fixed abundances 
that influence the others via the NV K and Ni L emission lines.
For instance, when we used the abundances for the north region (model 3), 
the N abundance changes from 0.3 to 0 solar and to compensate for the decreased 
photon counts in the 0.5--0.6 keV band, the normalization of the lower-temperature 
component ($kT\sim0.25$ keV) increases and the abundances of the other 
elements are suppressed.
When the abundances for the south region (model 4) are utilized,
the Ni L lines in the 0.8--1.4 keV range increases and the 
abundances of the other elements related to this energy band 
decreases.
The abundances of the blob region approaches to those 
of the $\eta$ Car region (figure \ref{fig:abd}~b) 
if we use these abundance sets.

The abundance values are therefore strongly influenced 
by the fixed abundances.
In model 2, all the abundances are around one solar, while
in model 1, 3 and 4, the best-fit metal abundances are far less 
than one.
This is due to the fact that the metal abundance and
emission measure are coupled with each other.
To decouple these parameters, we need more precise spectral 
measurements with X-ray microcalorimeters in future missions. 

In spite of the difficulty to determine the absolute abundances, 
abundance patterns of the blob region in models 1-4 are 
similar to those in the $\eta$ Car regions.
This similarity is another line of evidence 
that the diffuse X-ray emission in the eastern tip region and the 
$\eta$ Car region has the same origin.

Because the abundance patterns are rather independent of the fixed abundances, 
we can also evaluate the abundance ratio of different elements.
Importantly, there is no significant overabundance 
in nitrogen to oxygen that is expected from the optical, UV and X-ray
spectroscopy of $\eta$ Car and WR 25 \citep{Davidson1982, vanderHucht1981, Tsuboi1997}.
This result agrees with that in the $\eta$ Car region \citep{Hamaguchi2007} and
suggests that, if stellar winds produced the diffuse X-ray plasma,
the main drivers of the winds are main-sequence OB stars, and not
evolved massive stars.
Also the observed Fe/O ratio of 1.3--1.9 is too high 
compared to that of a type-II supernova, 0.5 \citep{Tsujimoto1995}. 
Although the Fe/O ratio increases up to 1 in type-II SNRs 
for less massive stars ($\sim13M_{\odot}$, table 1 in 
\cite{Tsujimoto1995}), this contradicts the fact that
the more massive stars evolve faster and explode earlier.

We note that similar situations exist in other massive star 
forming regions. For example, detailed spectral study of 
the diffuse X-ray emission has been conducted with Suzaku 
in M17 \citep{Hyodo2008}. 
We plot the abundance pattern of the diffuse
plasma in M17 in figure \ref{fig:abd}~c.
As is the case for the Carina Nebula, it shows a 
high Fe/O ratio and not enhanced N-to-O ratio, 
the latter of which may be natural because M17 contains 
no WR stars \citep{Townsley2003}.
Its subsolar abundances may be affected by the
fixed values at 0.3 solar.

Thus, both the stellar-wind and SNR interpretations are 
possible in terms of the absolute abundances.
The abundance pattern may favor plasma heating by winds from main-sequence OB stars.
Another possibility is a mixture of stellar winds and SNRs.
Also, since hot shocked gas by stellar winds and/or SNRs mix up 
interstellar gas, we may only see the interstellar abundances 
rather than stellar and/or SNR abundances.
Further studies are necessary from both observational 
and theoretical aspects.

\subsection{Hard X-ray component}
\label{sec:diss:hardX}

The X-ray spectrum above 2~keV can be 
explained by contribution of CXB, GRXE and known point sources.
This means that the residual hard X-ray emission,
seen around $\eta$~Carinae 
\citep{Hamaguchi2007}, localizes within $\sim$30$'$ from $\eta$~Car.
The scale size is consistent with an apparently extended hard X-ray emission
discovered with the GINGA satellite \citep{Koyama1990}
though the GINGA result did not count out emission from known point sources.
Because soft ($<$2~keV) diffuse X-ray spectra of the eastern tip region and the 
southern part of the $\eta$~Car region are almost identical,
the diffuse plasma would not relate to the residual hard X-ray emission.
Probably, the residual emission originates from a large number of low 
mass young stars embedded in the cloud around $\eta$~Car,
which are too faint to be detected individually.

\section{Conclusion}

In the present paper, we have investigated the properties of diffuse 
X-ray emission associated with the eastern tip region of the Carina 
Nebula using the Suzaku and XMM-Newton data.
Our conclusion is as follows.

(1) Strong extended X-ray emission was detected from the entire
field-of-view of Suzaku with a 0.2--5 keV surface brightness of 
0.7$\sim$4$\times10^{-14}$ erg s$^{-1}$ arcmin$^{-2}$.
Comparisons with the estimated contamination from astrophysical 
backgrounds and point sources suggest that most of the emission 
is diffuse in nature.

(2) The observed absorption column density and temperature are
consistent with those in the $\eta$Car region, suggesting the same origin as
the diffuse X-ray emission in the vicinity of the Carina Nebula.

(3) Estimated physical properties of the plasma such as pressure,
total energy, and mass can be explained by stellar winds from 
OB stars in the Carina Nebula or young SNR(s) with the age less 
than $\sim$10$^4$ yr. The SNR interpretation can provide the 
necessary energy and mass more easily.

(4) Absolute abundance values are strongly affected by abundances of 
metals fixed in spectral fits, allowing both the stellar-wind and 
SNR interpretations. The low nitrogen-to-oxygen and high 
iron-to-oxygen ratios derived from the spectral fits 
may support that the diffuse plasma heated up by stellar winds 
from main-sequence OB stars. The abundance ratios can be produced 
by a mixture of stellar winds and SNRs, as well.

The authors acknowledge discussion with Y. Hydo.
K.H. is supported by the NASA Astrobiology Program under RTOP 344-53-51.

\clearpage


\clearpage


\clearpage


\clearpage


\clearpage


\clearpage


\bigskip


\clearpage


\begin{table}[p]
  \caption{Properties of XMM-Newton point sources.}\label{tbl:srclist}
  \begin{center}
    \begin{tabular}{lcccccccc}

      \hline
      No. & R.A.$^{\rm a}$  & Decl.$^{\rm a}$ 
      & $C_{\rm MOS}$$^{\rm b}$ & $C_{\rm pn}$$^{\rm b}$
      & Flux$^{\rm c}$  & Var$^{\rm d}$ & Counterpart$^{\rm e}$ \\
      \hline
      
      1   & 10:47:31.68  & -59:32:33.7  
      & 199$\pm13$       & 411$\pm30$  
      &   3.8     &  No & 2MASS J 10473148-5932345       \\

      2	  & 10:49:28.08  & -59:34:55.2  
      &  68$\pm8$        & 124$\pm$17  
      &   1.1     &  No & 2MASS J 10492780-5935000       \\

      3$^{\rm f}$	  & 10:48:52.80  & -59:40:39.4  
      &  62$\pm9$        &  22$\pm13$   
      &   0.07    &  Yes & 2MASS J 10485248-5940392       \\ 

      4	  & 10:47:22.80  & -59:23:37.3  
      &   66$\pm8$       &  174$\pm$19
      &    0.38   &  No & 2MASS J 10472233-5923392       \\

      5	  & 10:46:45.84  & -59:30:13.0  
      &   30$\pm7$       &  83$\pm15$
      &    0.27   &  No & \begin{tabular}{c}
	                2MASS J 10464595-5930130       \\
			GSC 0862602325\\
                        \end{tabular}\\

      6	  & 10:47:05.28  & -59:30:27.0  
      &    15$\pm7$      &  61$\pm17$   
      &   0.20    &  No & 2MASS J 10470505-5930247       \\

      7$^{\rm g}  $& 10:46:09.12& -59:43:08.0  
      &    ---           &  336$\pm28$
      &   0.79   &  No & 2MASS J 10460901-5943025       \\

      8  & 10:48:10.80  & -59:41:44.9  
      &   47$\pm10$      &  73$\pm22$   
      &   0.23   &  No & 2MASS J 10480990-5941489       \\

      9  & 10:48:30.96  & -59:42:14.8  
      &   26$\pm8$       &   78$\pm19$   
      &   0.25   &  No & 2MASS J 10483064-5942144       \\

      10  & 10:48:14.16  & -59:43:32.2 
      &   15$\pm9$      &  81$\pm19$   
      &   0.26   &  No & ---      \\







        \hline
    \end{tabular}
  \end{center}
{\noindent
  $^{\rm a}$ Source positions in J2000 coordinates.\\
  $^{\rm b}$ Background-subtracted photon counts 
  detected with EPIC MOS and pn in 0.4--10 keV. 
  $C_{\rm MOS}$ is the average counts of MOS1 and MOS2. 
  Errors are 1$\sigma$.\\
  $^{\rm c}$ The 0.4--10 keV X-ray flux in 10$^{-13}$ erg s$^{-1}$ cm$^{-2}$ 
  which corresponds to 6$\times10^{31}$ erg s$^{-1}$ at 2.3 kpc.
  Fluxes of No.1, 2, 4, and 7 are derived from the spectral fittings,
  while the other are estimated from $C_{\rm pn}$ assuming the thin-thermal
  plasma model (see text).\\
  $^{\rm d}$ Time variability based on the $\chi^2$ statistics.\\ 
  $^{\rm e}$ Counterpart candidates within 10 arcsec of the X-ray position,
  based on searches of the 2MASS and AGAST catalogs.\\
  $^{\rm f}$ This source falls in the CCD gap of pn.\\
  $^{\rm g}$ This source is outside the FOV of MOS.\\
}
\end{table}


\begin{table}[p]
  \caption{Results of the spectral fits to the point sources$^{\rm a}$.}
  \label{tbl:srcfit}

  \begin{center}
    \begin{tabular}{lccccc}

      \hline
      No. & $N_{\rm H}$$^{\rm b}$ 
      & $kT$$^{\rm c}$ 
      & Normalization$^{\rm d}$ 
      & $L_{\rm X}$$^{\rm e}$ 
      & $\chi^2$/d.o.f. \\ \hline
      
      1 & 0.17$_{-0.12}^{+0.57}$    
      & 0.42$_{-0.18}^{+0.45}$, $>32$
      & 4.8$_{-2.9}^{+4.1}$$\times10^{-5}$, 2.5$_{-0.6}^{+0.5}$$\times10^{-4}$
      & 3
      & 19.5/25 \\

      2 & 9.6$_{-3.8}^{+6.3}$    
      & 1.8$_{-0.8}^{+2.3}$
      & 1.0$\pm{0.1}$$\times10^{-3}$ 
      & 6
      & 3.5/7 \\

      4 & $<0.95$
      & 0.70$_{-0.44}^{+0.10}$
      & 4.4$\pm0.7$$\times10^{-5}$ 
      & 0.3
      & 7.1/9 \\

      7 & $0.72_{-0.16}^{+0.17}$ 
      & $0.12_{-0.03}^{+0.04}$ 
      & 8.4$_{-4.5}^{+4.6}$$\times10^{-2}$  
      & 80
      & 37.6/23 \\

        \hline
    \end{tabular}
  \end{center}
{\noindent

  $^{\rm a}$ A single plasma model is assumed for No. 2, 4, and 7, while a two temperature 
  model is used for No.1. A metal abundance is fixed at 0.3 solar value. 
  Errors refer to the 90\% confidence range. \\
  $^{\rm b}$ Hydrogen column density in 10$^{22}$ cm$^{-2}$.\\
  $^{\rm c}$ Plasma temperature in keV.\\
  $^{\rm d}$ Normalization factor of the APEC model, representing 
  10$^{-14}$/4$\pi D^2$ $EM$, where $D$ is
  a distance to the Carina nebula and $EM$ is an emission measure in 
  cm$^{-3}$.\\
  $^{\rm e}$ Absorption-corrected 0.4--10 keV luminosity 
  in 10$^{32}$ erg s$^{-1}$ assuming a distance of 2.3 kpc.\\

}
\end{table}


\begin{table}[p]
  \caption{Results of the two-temperature plasma model fit to the diffuse X-ray emission in the blob region.}
  \label{tbl:fit1}

  \begin{center}
    \begin{tabular}{lccccc}

\hline\hline
Model$^{\rm a}$                               &  1           &   2          & 3            & 4            & Typical error$^{\rm b}$ \\
\hline									       	        								   
									       	        								   
Two-temperature plasma component$^{\rm c}$\\				       	        			   
									       	        								   
$N_{\rm H}$ (10$^{22}$ cm$^{-2}$)              & 0.23         & 0.25         & 0.22         & 0.26         & 0.01\\ 
$kT_1$  (keV)                                & 0.25         & 0.24         & 0.25         & 0.24         & 0.01  \\ 
$kT_2$  (keV)                                & 0.55         & 0.56         & 0.55         & 0.54         & 0.01  \\ 
C       (solar)                              & 0.3 (fixed)  & 1.0 (fixed)  & 0.0 (fixed)  & 0.0 (fixed)  & $-$   \\
N       (solar)                              & 0.3 (fixed)  & 1.0 (fixed)  & 0.0 (fixed)  & 0.0 (fixed)  & $-$   \\
O       (solar)                              & 0.24         & 0.55         & 0.16         & 0.10         & 0.01  \\ 
Ne      (solar)                              & 0.46         & 0.93         & 0.36         & 0.21         & 0.01  \\ 
Mg      (solar)                              & 0.44         & 0.94         & 0.32         & 0.24         & 0.01  \\ 
Al      (solar)                              & 0.3 (fixed)  & 1.0   (fixed)& 0.075 (fixed)& 0.029 (fixed)& $-$   \\
Si      (solar)                              & 0.54         & 1.1          & 0.40         & 0.37         & 0.02  \\ 
S       (solar)                              & 0.74         & 1.5          & 0.56         & 0.57         & 0.1   \\ 
Ar      (solar)                              & 0.3 (fixed)  & 1.0 (fixed)  & 0.0 (fixed)  & 0.13 (fixed) & $-$   \\ 
Ca      (solar)                              & 0.3 (fixed)  & 1.0 (fixed)  & 0.0 (fixed)  & 0.0 (fixed)  & $-$   \\ 
Fe      (solar)                              & 0.32         & 0.71         & 0.23         & 0.19         & 0.01  \\ 
Ni      (solar)                              & 0.3 (fixed)  & 1.0 (fixed)  & 0.089 (fixed)& 0.78 (fixed) & $-$   \\ 
log$EM_1$ (cm$^{-3}$ arcmin$^{-2}$)           & 54.9         & 54.6         & 55.0         & 55.4         & 0.02 \\ 
log$EM_2$ (cm$^{-3}$ arcmin$^{-2}$)           & 54.5         & 54.2         & 54.6         & 54.7         & 0.02 \\ 
Flux1 (10$^{-14}$ erg s$^{-1}$ arcmin$^{-2}$)  & 1.8          & 1.8          & 1.8          & 2.3          & 0.1 \\ 
Flux2 (10$^{-14}$ erg s$^{-1}$ arcmin$^{-2}$)  & 2.2          & 2.2          & 2.2          & 1.8          & 0.1 \\ 
\hline									   									   
									   								   
Power-law component$^{\rm d}$        \\  				   				   
Flux (10$^{-14}$ erg s$^{-1}$ arcmin$^{-2}$) & 0.56         & 0.57         & 0.56         & 0.56         & 0.02  \\ 
\hline									   									   
									   									   
LHB component$^{\rm e}$              \\					   					   
Flux (10$^{-14}$ erg s$^{-1}$ arcmin$^{-2}$) & 0.10         &  0.11        &  0.10        &  0.10        & 0.02  \\ 
\hline									   									   
									   									   
$\chi^2$/d.o.f.                              &  1.24        & 1.31         & 1.21         & 1.22         &       \\
         d.o.f.                              &  1231        & 1231         & 1231         & 1231         &       \\
\hline


    \end{tabular}

  \end{center}
{\noindent

  $^{\rm a}$ Fitting models with different fixed abundances.  \\
  $^{\rm b}$ Typical fitting errors at the 90\% confidence level.\\ 
  $^{\rm c}$ A commonly-absorbed plasma model.
  Arabic numbers 1 and 2 denote the two temperature components.
  Parameter definitions are the same as those in table \ref{tbl:srcfit}.
  Fluxes are calculated in 0.2--5 keV.\\
  $^{\rm d}$ A power-law model representing CXB, GRXE, and point sources. 
  A photon index is fixed at 1.5. The same
  absorption for the two-temperature plasma is assumed. Normalization
  is photon keV$^{-1}$ cm$^{-2}$ at 1 keV.\\
  $^{\rm e}$ A single-temperature plasma model representing LHB.
  A plasma temperature is fixed at 0.1 keV.\\

}
\end{table}


\begin{table}[p]
  \caption{Results of the two-temperature plasma model fit to the diffuse X-ray emission in the east and nw regions$^{\rm a}$.}
  \label{tbl:fit2}

  \begin{center}
    \begin{tabular}{lccccc}

\hline\hline
Region                                         &  east                  &   nw                      \\
\hline

Two-temperature plasma component\\

$N_{\rm H}$ (10$^{22}$ cm$^{-2}$)              & 0.21$_{-0.07}^{+0.09}$ & 0.32$_{-0.07}^{+0.11}$     \\ 
$kT_1$  (keV)                                  & 0.20$_{-0.02}^{+0.04}$ & 0.19$_{-0.03}^{+0.02}$     \\ 
$kT_2$  (keV)                                  & 0.54$_{-0.07}^{+0.04}$ & 0.41$_{-0.08}^{+0.10}$     \\ 
C       (solar)                                & 0.3 (fixed)            & 0.3 (fixed)                \\ 	    
N       (solar)                                & 0.3 (fixed)            & 0.3 (fixed)                \\ 	    
O       (solar)                                & 0.15$_{-0.06}^{+0.12}$ & 0.07$_{-0.02}^{+0.04}$     \\ 
Ne      (solar)                                & 0.33$_{-0.14}^{+0.31}$ & 0.27$_{-0.05}^{+0.14}$     \\ 
Mg      (solar)                                & 0.30$_{-0.14}^{+0.29}$ & 0.25$_{-0.10}^{+0.16}$     \\ 
Al      (solar)                                & 0.3 (fixed)            & 0.3 (fixed)                \\   	    
Si      (solar)                                & 0.43$_{-0.20}^{+0.37}$ & 0.96$_{-0.43}^{+0.57}$     \\ 
S       (solar)                                & 0.74 (fixed)           & 0.74 (fixed)               \\ 
Ar      (solar)                                & 0.3 (fixed)            & 0.3 (fixed)                \\ 	    
Ca      (solar)                                & 0.3 (fixed)            & 0.3 (fixed)                \\ 	    
Fe      (solar)                                & 0.16$_{-0.05}^{+0.10}$ & 0.17$_{-0.06}^{+0.03}$     \\ 
Ni      (solar)                                & 0.3 (fixed)            & 0.3 (fixed)                \\   	    
log$EM_1$ (cm$^{-3}$ arcmin$^{-2}$)            & 54.3$_{-0.5}^{+0.4}$   & 55.0$\pm{0.7}$       \\ 
log$EM_2$ (cm$^{-3}$ arcmin$^{-2}$)            & 54.0$\pm{0.2}$         & 54.3$_{-0.3}^{+0.8}$       \\ 
Flux1 (10$^{-14}$ erg s$^{-1}$ arcmin$^{-2}$)  & 0.23$_{-0.15}^{+0.41}$ & 0.41$\pm{0.08}$     \\ 
Flux2 (10$^{-14}$ erg s$^{-1}$ arcmin$^{-2}$)  & 0.48$_{-0.20}^{+0.31}$ & 0.45$_{-0.20}^{+0.11}$     \\ 
\hline
	
Power-law component        \\    
Flux  (10$^{-14}$ erg s$^{-1}$ arcmin$^{-2}$)  & 0.56$\pm{0.06}$          & 0.55$\pm{0.09}$          \\ 
\hline

LHB component              \\
Flux  (10$^{-14}$ erg s$^{-1}$ arcmin$^{-2}$)  & 0.087$_{-0.032}^{+0.031}$& 0.15$_{-0.06}^{+0.05}$  \\ 
\hline

$\chi^2$/d.o.f.                                & 0.77                     & 0.60                    \\
         d.o.f.                                & 245                      & 111                     \\
\hline
    \end{tabular}

  \end{center}
{\noindent

  $^{\rm a}$ Notations and symbols are the same as table \ref{tbl:fit1}.

}
\end{table}

\clearpage

\begin{table}[p]
  \caption{Result of the two-temperature plasma model fit to the XMM MOS spectra of the blob region$^{\rm a}$.}
  \label{tbl:fit3}

  \begin{center}
    \begin{tabular}{lccccc}

\hline\hline
Model                                         &  1                   \\
\hline

Two-temperature plasma component\\

$N_{\rm H}$ (10$^{22}$ cm$^{-2}$)             & 0.19$_{-0.02}^{+0.03}$  \\ 
$kT_1$  (keV)                                                    & 0.24$\pm{0.01}$              \\ 
$kT_2$  (keV)                                                    & 0.58$\pm{0.01}$              \\ 

C       (solar)                                                        & 1.2$_{-0.9}^{+0.4}$            \\
N       (solar)                                                        & 0.3 (fixed)                          \\ 	    
O       (solar)                                  & 0.23$_{-0.03}^{+0.02}$      \\ 
Ne      (solar)                                 & 0.44$\pm{0.07}$                  \\ 
Mg      (solar)                                & 0.46$\pm{0.07}$                   \\ 
Al      (solar)                                  & 0.3 (fixed)                               \\   	    
Si      (solar)                                  & 0.48$_{-0.08}^{+0.07}$      \\ 
S       (solar)                                  & 0.48$_{-0.18}^{+0.20}$      \\ 
Ar      (solar)                                  & 0.3 (fixed)                              \\ 	    
Ca      (solar)                                & 0.3 (fixed)                               \\ 	    
Fe      (solar)                                 & 0.32$_{-0.04}^{+0.05}$        \\ 
Ni      (solar)                                 & 0.3 (fixed)                                 \\   	    
log$EM_1$ (cm$^{-3}$ arcmin$^{-2}$)               & 54.8$\pm{0.1}$                    \\ 
log$EM_2$ (cm$^{-3}$ arcmin$^{-2}$)               & 54.5$\pm{0.1}$                    \\ 
Flux1 (10$^{-14}$ erg s$^{-1}$ arcmin$^{-2}$)  & 1.7$\pm{0.2}$               \\ 
Flux2 (10$^{-14}$ erg s$^{-1}$ arcmin$^{-2}$)  & 2.7$_{-0.4}^{+0.2}$     \\ 
\hline
	
Power-law component        \\    
Flux  (10$^{-14}$ erg s$^{-1}$ arcmin$^{-2}$)   & 0.84$\pm{0.05}$                    \\ 
\hline

LHB component              \\
Flux  (10$^{-14}$ erg s$^{-1}$ arcmin$^{-2}$)  & $<0.06$  \\ 
\hline

$\chi^2$/d.o.f.                                & 1.20                       \\
         d.o.f.                                       & 337                        \\
\hline
    \end{tabular}

  \end{center}
{\noindent

  $^{\rm a}$ Notations and symbols are the same as table \ref{tbl:fit1}.

}
\end{table}

\clearpage


\begin{table}[p]
  \caption{Physical properties of the diffuse plasma in the blob region$^a$.}
  \label{tbl:plasma}

  \begin{center}
    \begin{tabular}{lccccc}

\hline\hline
Parameter                                    &  Scale Factor                &       $T_1$                     &      $T_2$\\
\hline

\multicolumn{4}{c}{Observed X-ray Properties}\\
\hline

$kT$ (keV)                                    &   $-$                               &        0.3                          &      0.6 \\
$L_{\rm X}$ (ergs s$^{-1}$)       &   $-$                               &   2$\times10^{34}$     &     1$\times10^{34}$\\
$V$ (cm$^{3}$)                            &   $\eta$                          &   1$\times10^{57}$     &     1$\times10^{57}$\\
\hline

\multicolumn{4}{c}{Estimated X-ray Plasma Properties}\\
\hline


$n_{\rm e}$ (cm$^{-3}$)             &   $\eta^{-1/2}$            &        0.3                          &      0.4 \\
$P/k$ (K cm$^{-3}$)                    &   $\eta^{-1/2}$            &   2$\times10^{6}$       &     5$\times10^{6}$\\
$U$ (ergs)                                     &   $\eta^{1/2}$             &   4$\times10^{47}$     &     1$\times10^{48}$\\
$t_{\rm cool}$ (yr)                         &  $\eta^{1/2}$             &   1$\times10^{6}$       &     4$\times10^{6}$\\
$M_{\rm plasma}$ ($M_\odot$) &  $\eta^{1/2}$             &   0.2                               &     0.2 \\

\hline

    \end{tabular}

  \end{center}
{\noindent

  $^{\rm a}$ $\eta$ is a filling factor for the volume of the 
             plasma. $T_1$ and $T_2$ indicate the two temperature plasma
             component in table \ref{tbl:fit1} model 1.

}
\end{table}

\end{document}